\documentclass[floats,aps,showpacs,amssymb,prl,twocolumn,endnotes,groupedaddress,superscriptaddress
,secnumarabic%
,tightenlines%
]{revtex4}
\usepackage{amssymb}
\usepackage{graphics}
\usepackage{amsmath}
\usepackage{amsfonts}
\usepackage{bm}
\usepackage{graphicx,amssymb,amsmath}

\def\lf{\left}

\def\non{\nonumber}

\def\ri{\right}
\def\al{\alpha}

\def\1{{_{1}}}\def\2{{_{2}}}

\def\noHe0{:\;\!\!\;\!\!:H_e(0):\;\!\!\;\!\!:}
\def\noHm0{:\;\!\!\;\!\!:H_\mu(0):\;\!\!\;\!\!:}

\def\lf{\left}

\def\non{\nonumber}

\def\ri{\right}

\def\al{\alpha}

\def\1{{_{1}}}\def\2{{_{2}}}

\begin{document}

\title{Decoherence in neutrino oscillations,  neutrino nature and $CPT$ violation}

\author{A. Capolupo }
\affiliation{Dipartimento di Fisica "E.R. Caianiello" Universit\'a di Salerno,  and INFN - Gruppo Collegato di Salerno, Italy.}
\author{ S.M. Giampaolo }
\affiliation{Division of Theoretical Physics,  Ruder Bo\v{s}kovi\'{c} Institute, Bijen\u{c}ka cesta 54, 10000 Zagreb, Croatia.}
\author{G. Lambiase}
\affiliation{Dipartimento di Fisica "E.R. Caianiello" Universit\'a di Salerno,  and INFN - Gruppo Collegato di Salerno, Italy.}

\date{\today}
\def\be{\begin{equation}}
\def\ee{\end{equation}}
\def\al{\alpha}
\def\bea{\begin{eqnarray}}
\def\eea{\end{eqnarray}}

\begin{abstract}

We analyze many aspects of the phenomenon of the  decoherence for neutrinos propagating in  long baseline experiments.
We show that, in the presence of an off-diagonal term in the dissipative matrix, the Majorana neutrino can violate the $CPT$ symmetry, which, on the contrary, is preserved for Dirac neutrinos.
 We show that oscillation formulas for Majorana neutrinos depend  on the choice of the mixing matrix $U$. Indeed, different choices of $U$ lead to different oscillation formulas.
Moreover, we study the possibility to reveal the differences between Dirac and Majorana neutrinos in the oscillations.
We use the present values of the experimental parameters in order to relate our theoretical proposal with experiments.

\end{abstract}


\maketitle

\section{Introduction }

The phenomena of neutrino  mixing and oscillations, induced by the non-zero neutrino mass,
represent an hint of physics beyond the Standard Model of particles.
 It has been confirmed by many experiments
 \cite{Nakamura1}-\cite{Nakamura6}. At the present, the main issues of the neutrino physics are the determination of the absolute neutrino mass and its nature.  As a matter of fact, since neutrino is electrically
neutral, two possibilities exist, either neutrino is distinct from its antiparticle  and hence is of the Dirac
type, or it is equal to its antiparticle  and it is of the  Majorana type.

 To reveal the neutrino nature, many experiments, based on the detection of the neutrinoless double beta decay, have been proposed  \cite{Giuliani}. Recently, it has been shown that quantities such as the Leggett-Garg $K_{3}$ quantity \cite{Richter} and  the geometric phase for neutrinos  \cite{Capolupo2018}, can, in principle, discriminates between Dirac and Majorana neutrinos. Moreover, it has been shown that in the presence of decoherence,  the neutrino oscillation formulas can depend on the Majorana phase \cite{Benatti}.
However, at the moment the nature of the neutrino remains an open question.

On the other hand, particle mixing phenomenon, in particular the $B^0 - \overline{B}^0$ mixing is used to test the $CPT$ symmetry. The $CPT$ theorem, affirms that the simultaneous transformations of charge conjugation $C$, parity transformation $P$, and time reversal $T$, is  an exact symmetry of nature at the fundamental level \cite {Kost}.

In this paper, we show that, if quantum decoherence appears in neutrino oscillations,
then  long baseline experiments   might allow to investigate the  nature of neutrinos and the $CPT$ symmetry.
Moreover, if the neutrinos are Majorana particles, the decoherence could allow the right choice of the matrix mixing $U$.

The phenomena of dissipation and decoherence are consequences of  interaction with the environment, which, in neutrino case, could be originated by quantum gravity effects, or strings and branes.
A significant research effort had been undertaken in
the study of dissipation in neutrino oscillations \cite{Benatti,BenattiB,Benatti1}.
It has been shown that such phenomena can modify the oscillation frequencies and the oscillation formulas  \cite{Benatti,BenattiB,Benatti1}.
Moreover, it has been  noted that   the dissipation can generate oscillation formulas for Majorana neutrinos different with respect to the ones for Dirac neutrinos \cite{Benatti}.
Still, other theoretical results can be obtained which are extremely relevant.

Here, by  considering the neutrino as an open quantum system  interacting with its environment,
 we analyze many aspects of the decoherence effect in flavor mixing.
We study the time evolution of the density matrix representing the neutrino state in the flavor basis and we analyze  the case in which the matrix describing the dissipator has off-diagonal terms.
Specifically, we reveal the possible $CPT$ symmetry breaking in the Majorana neutrino oscillation,
and we study  the differences between Majorana and Dirac neutrinos.
We  prove that,   the presence of  off-diagonal terms in decoherence matrix, leads to probability of transitions depending on the representation of the Majorana mixing matrix. Thus, if the decoherence exists in neutrino propagation, the oscillation formulas could provide a tool to determine the choice of the mixing matrix U.
Moreover, by considering the data of IceCube and DUNE experiments and the recent constraints on decoherence parameters \cite{Coloma,Balieiro},
we show that long baseline experiments on atmospheric neutrinos,   like IceCube experiment, could reveal the nature of neutrino and could allow to test the $CPT$ symmetry.

We analyze the neutrinos propagation in the vacuum and  through a  medium.
The matter effects, are taken into account by replacing in the oscillation formulas in vacuum, $\Delta m^{2}$ with $\Delta m_{m}^{2} = \Delta m^{2} R_{\pm}$, and $\sin 2\theta $ with $\sin 2\theta_{m} =  \sin 2\theta /R_{\pm}$.
The coefficients $R_{\pm}$ describing the Mikhaev-Smirnow-Wolfenstein (MSW) effect \cite{MSW1,MSW2} are given by, $R_{\pm} = \sqrt{\left(\cos 2\theta  \pm  \frac{2 \sqrt{2 } G_{F} n_{e} E}{\Delta m^{2}}\right)^{2}+ \sin^{2} 2\theta}\,, $  with $+$ for oscillation of antineutrinos and $-$ for oscillations of neutrinos.
%
Here, $n_{e }$ is the electron density in the matter, $G_F$ is the Fermi constant, and $E$ is the neutrino energy.
Notice that the MSW effect can be  relevant in the $\nu_{e}\leftrightarrow\nu_{\mu}$ oscillations, since the $\nu_{e}$ and $ \nu_{\mu}$  indices of refraction are different in media like the Earth and the Sun, $\kappa(\nu_{e}) \neq \kappa(\nu_{\mu})$. In particular, $\kappa(\nu_{e}) - \kappa(\nu_{\mu}) = - \sqrt{2 } G_{F} n_{e} / p$. On the contrary, in the case of the $\nu_{\mu } \leftrightarrow \nu_{\tau}$ mixing,  the $\nu_{\mu}$ and $ \nu_{\tau}$  indices of refraction are different only in very dense matter, like the core of supernovae, but they are almost identical in the matter of  Earth and the Sun. Therefore, in such media, $R \sim 1$ and the $\nu_{\mu} \leftrightarrow \nu_{\tau}$ oscillations are almost identical to  the ones in vacuum \cite{MSW3}.
In the following we consider  the propagation through  Earth  in which the MSW effect is relevant only for $\nu_{e}\leftrightarrow\nu_{\mu}$ oscillations.

The paper is organized as follows. In Sec.II we report the main differences between Majorana and Dirac neutrinos.  In Sec.III we analyze quantum decoherence in neutrino propagation and we show the effects induced on neutrino   by an off-diagonal term in the dissipation matrix.  Numerical analysis, for neutrino propagation in vacuum and through matter, are reported in Sec.IV. Sec.V is devoted to the conclusions.

\section{Majorana and Dirac neutrinos}

A very important difference between Dirac and Majorana neutrinos consists in the fact that, while the Dirac Lagrangian is invariant under $U(1)$ global transformation, and hence the charges associated  (electric, leptonic, etc.) with the transformations are conserved, the Majorana Lagrangian breaks the $U(1)$  symmetry \cite{note}.
 This fact implies that processes violating
 the lepton number, such as neutrinoless double beta decay, are allowed for Majorana neutrinos and forbidden for Dirac ones.
In the case of neutrino mixing, the breaking of the $U(1)$ global symmetry of the Majorana Lagrangian implies that
the Pontecorvo-Maki-Nakagawa-Sakata (PMNS) mixing matrix with dimension  $n \times n $,
contains a total number of physical phases  for Majorana neutrinos  different with respect to the one of Dirac neutrinos. Indeed, in the case of the mixing of $n$ Dirac fields, one has $N_{D}$ physical phases given by $N_{D} = \frac{(n -1)(n-2)}{2}$, and in the case of the mixing of $n$  Majorana fields, one has $N_{M}$ phases given by  $N_{M} = \frac{n (n - 1) }{2}$. The $n -1$ extra phases present in the Majorana neutrino mixing (called Majorana phases) represent another important distinction between Dirac and Majorana neutrinos. The detection of such phases can allow to fix the nature of neutrinos.

  The mixing matrices for Majorana $U_M$ and for Dirac neutrinos $U_D$ can be related by the equation,
\bea
U_M = U_D \cdot diag(1, e^{i \phi_1}, e^{i \phi_2},..., e^{i \phi_{n-1}} )\,,
\eea
where $\phi_i$, with $i = 1,...,n-1,$ are the Majorana phases.
 Other representations of  Majorana mixing matrix can be  obtained by the rephasing the lepton charge fields in the charged current weak-interaction Lagrangian, (for details see Ref.\cite{Giunti}).
For example, for two mixed Majorana fields,
one can consider the following mixing relations
  \bea\label{U1}
\left(
  \begin{array}{c}
    \nu_{e} \\
    \nu_{\mu} \\
  \end{array}
\right) =
   \left(
            \begin{array}{cc}
             \cos  \theta   &    \sin  \theta\,  e^{i \phi}\\
             -  \sin  \theta  &  \cos  \theta\,  e^{i \phi}\\
            \end{array}
          \right) \left(
  \begin{array}{c}
    \nu_{1} \\
    \nu_{2} \\
  \end{array}
\right)
           \,,
          \eea
          or
          \bea\label{U2}
\left(
  \begin{array}{c}
    \nu_{e} \\
    \nu_{\mu} \\
  \end{array}
\right) =    \left(
            \begin{array}{cc}
             \cos  \theta   &    \sin  \theta\,  e^{-i \phi}\\
             -  \sin  \theta\,  e^{i \phi} &  \cos  \theta \\
            \end{array}
          \right)\left(
  \begin{array}{c}
    \nu_{1} \\
    \nu_{2} \\
  \end{array}
\right)
           \,,
          \eea
where $\theta$ is the mixing angle and $\phi$ is the Majorana phase.
Such a phase can  be removed for Dirac neutrinos  by rephasing the mass term of the Dirac Lagrangian.
  For example, in  Eq.(\ref{U2}),  $\phi$ is eliminated  by means of the replacements, $\nu_{1} \rightarrow \nu_{1}$ and $\nu_{2} \rightarrow \nu_{2}\, e^{ i \phi}$.

Notice that, in the case of absence of decoherence (i.e. in standard treatment of neutrino mixing) and in the case of dissipation with diagonal decoherence matrix,
the probabilities of neutrino transitions are invariant under the rephasing $\textit{U}_{\alpha k} \rightarrow e^{i \phi_{k}} \textit{U}_{\alpha k} $  $(\alpha = e, \mu; k =1,2)$. Then, in such cases, the presence of the Majorana phases $\phi_{i}$ do  not affect the oscillation formulas for neutrino propagating in the vacuum, through matter and through a magnetic field, being such formulas equivalent for Majorana and for Dirac neutrinos \cite{Pontecorvo}.
On the contrary,  in the presence of off-diagonal terms in decoherence matrix, the oscillation formulas for Majorana neutrinos, depend on the phases $\phi_{i}$  \cite{Benatti}.

 Moreover,   we will reveal other two important aspects:
  a) Majorana neutrinos can violate $CPT$ symmetry;
  b)  different choices of the mixing matrix for such neutrinos can lead to different probability of transitions, (for example, the oscillation formulas obtained by using Eq.(\ref{U1}) can be different with respect to the ones obtained by means of Eq.(\ref{U2}), see below).

\section{Decoherence and neutrino oscillations}

The evolution of the neutrino considered as an open system, can be expressed by the Lidbland--Kossakowski master equation  \cite{Lind}
 \bea\label{evolution}
\frac{\partial \rho(t)}{\partial t} & =& -\frac{i}{\hbar}[H_{eff},\rho(t)] + D[\rho(t)]\,.
\eea
Here, $H_{eff} = H^{\dag}_{eff}$ is the effective hamiltonian, and $D[\rho(t)]$ is the dissipator defined as
\bea\label{D}
D[\rho(t)] = \frac{1}{2}\sum_{i,j=0}^{N^{2}-1} a_{i j } \left([F_{i} \,\rho(t)\,, F^{\dag}_{j}] + [F_{i}\,,\rho(t)\,F^{\dag}_{j}]  \right).
\eea
The coefficients $a_{i j } $ of the Kossakowski matrix,  could be derived  by the properties of the environment  \cite{Benatti} and
$F_i$, with $i = N^{2}-1$, are a set of operators such  that $Tr (F_{k}) = 0$ for any $k$ and $Tr (F^{\dag}_{i}F_{j}) = \delta_{ij}$.
In the three flavor neutrino mixing case, $F_{i}$ are represented by the Gell-Mann matrices $\lambda_{i}$. In the two flavor neutrino mixing $F_{i}$, are the Pauli matrices $\sigma_{i}$.

For the sake of simplicity, in the following we consider the mixing between two flavors  (our results can be extended to the three flavor neutrino mixing case).
Expanding Eqs.(\ref{evolution}) and (\ref{D}) in the bases of the $SU(2)$, Eq.(\ref{evolution}) can be written as
\bea\label{evol}
\frac{d \rho_{\lambda}}{d t} \sigma_{\lambda} = 2\, \epsilon_{ijk}\, H_{i} \,\rho_{j}(t)\, \sigma_{\lambda}\, \delta_{\lambda k} \,+ \, D_{\lambda \mu}\, \rho_{\mu}(t)\, \sigma_{\lambda}\,,
\eea
where $\rho_{\mu} = \mathrm{ Tr}(\rho \, \sigma_{\mu}) $, with $\mu \in [0, 3]$ and $D_{\lambda \mu}$ is a $4 \times 4$ matrix. The conservation of the probability implies $D_{\lambda 0} = D_{0 \mu} = 0$, then
\bea\label{Dissipator}
D_{\lambda \mu}\,=\,-
\left(
  \begin{array}{cccc}
    0 & 0 & 0 & 0 \\
    0 & \gamma_{1} & \alpha & \beta \\
    0 & \alpha & \gamma_{2} & \delta \\
    0 & \beta & \delta & \gamma_{3} \\
  \end{array}
\right).
\eea
All the parameters of Eq.(\ref{Dissipator}) are reals and the diagonal elements are positive in order to satisfy the condition, $Tr(\rho(t))=1\,$ for any time $t$.

In  order to study the effects of  a non-diagonal form of the decoherence matrix on neutrino physics,
we consider, for simplicity $D_{\lambda\mu}$ given by
 \begin{equation}\label{Dissipator2}
   D_{\lambda\mu}= -\left( \begin{array}{cccc}
   0 & 0 & 0 & 0 \\
   0 & \gamma & \alpha & 0 \\
   0 & \alpha & \gamma &  0 \\
   0 & 0 & 0 & \gamma_3 \end{array} \right)\,.
 \end{equation}
Such a dissipator is obtained by Eq.(\ref{Dissipator}), by
 setting, $  \gamma_1=\gamma_2=\gamma\,,  $
 and $\beta  = \delta = 0$.
  The condition of complete  positivity of the density matrix $\rho(t)$, implies $\forall t$,
   the following  condition  $|\alpha| \leq \gamma_{3}/2 \leq \gamma$.
  %
%
%
%
%
By setting, $\Delta =\frac{\Delta m^{2}}{2 E} $, and by taking into account Eq.(\ref{Dissipator2}),
one has $ \dot{\rho}_{0}(t) = 0 $, which for two neutrino families implies $ {\rho}_{0}(t) = 1/2$.
Then the master equation (\ref{evol}) can be written as
\bea\label{ddiag}
\left(
  \begin{array}{c}
    \dot{\rho}_{1}(t) \\
    \dot{\rho}_{2}(t) \\
    \dot{\rho}_{3}(t) \\
  \end{array}
\right)
= \left(
    \begin{array}{ccc}
     - \gamma & - \Delta + \alpha & 0 \\
      \Delta + \alpha & - \gamma & 0 \\
      0 & 0 & - \gamma_{3} \\
    \end{array}
  \right)
  \left(
  \begin{array}{c}
     {\rho}_{1}(t) \\
     {\rho}_{2}(t) \\
    {\rho}_{3}(t) \\
  \end{array}
\right)\,.
\eea
By solving Eq.(\ref{ddiag}), one gets
\bea\non\label{rho}
{\rho}_{1}(t) & = & e^{- \gamma t} \Big[ \rho_{1}(0)   \cosh  \lf(\Omega_{\alpha} t \ri)
+
 \rho_{2}(0)\, \frac{\sinh \lf(\Omega_{\alpha} t  \ri)}{\Omega_{\alpha}   }\, \Xi_{+}\Big]
\\\non
\\\non
{\rho}_{2}(t) & = & e^{- \gamma t} \Big[ \rho_{1}(0)\, \frac{\sinh \lf(\Omega_{\alpha} t  \ri)}{\Omega_{\alpha}   }\, \Xi_{-}
+
 \rho_{2}(0) \cosh  \lf(\Omega_{\alpha} t \ri)
\Big]
\\\non
\\
{\rho}_{3}(t) & = & \rho_{3}(0) e^{- \gamma_{3} t}\,,
\eea
where $\Xi_{\pm} = \alpha \pm \Delta$ and  $\Omega_{\alpha} = \sqrt{\alpha^{2} - \Delta^{2}}$.
%
%
Hence, the matrix density,  at any time, $t$ reads
\bea\label{matrX}
\rho(t) = \left(
            \begin{array}{cc}
              {\rho}_{0}(t) + {\rho}_{3}(t) &  {\rho}_{1}(t) - i {\rho}_{2}(t) \\
              {\rho}_{1}(t) + i {\rho}_{2}(t) & {\rho}_{0}(t) - {\rho}_{3}(t) \\
            \end{array}
          \right)\,.
\eea
%

By using the mixing relations for Majorana neutrinos, given in Eq.(\ref{U2}),
the matrix density of the electron neutrino, at time $t = 0$, is
\bea\label{Re0}
\rho_{e}(0) = \left(
  \begin{array}{cc}
    \cos^{2} \theta & \frac{1}{2} \sin 2 \theta \, e^{ i \phi} \\
   \frac{1}{2} \sin 2 \theta \, e^{- i \phi}  & \sin^{2} \theta  \\
  \end{array}
\right)\,,
\eea
and similar for muon neutrino.
Then at time $t$, one has
\bea\label{matrE}
\rho_{e}(t)  = \left(
                 \begin{array}{cc}
                  \Lambda_{+} &  \Theta^{*} \\
                    \Theta & \Lambda_{-} \\
                 \end{array}
               \right)\,,
\eea
where,
$
\Lambda_{\pm} = \frac{1}{2}\lf[1 \pm \cos 2 \theta e^{-\gamma_{3} t} \ri]\,,
$ and
\bea\non
\Theta =  \frac{\sin 2 \theta \, e^{-  \gamma   t - i \phi}}{2 \Omega_{\alpha} t} \Big\{\Omega_{\alpha} \cosh \lf(\Omega_{\alpha} t \ri)
+ i \Upsilon_{\alpha,\phi}\,\sinh \lf(\Omega_{\alpha} t  \ri) \Big\}\,,
\eea
with $\Upsilon_{\alpha,\phi} = e^{2 i \phi }\alpha - \Delta$.

The  probabilities of transition  $P_{\nu_{\sigma} \rightarrow \nu_{\varrho}}(t)$, with $\sigma $ and $\varrho$ neutrino flavors, are given by
$P_{\nu_{\sigma} \rightarrow \nu_{\varrho}}(t) = Tr \lf[\rho_{\varrho}(t) \rho_{\sigma}(0) \ri]$.
Explicitly, one has
\bea\non\label{pA}
&& P_{\nu_{\sigma} \rightarrow \nu_{\varrho}}(t) = \frac{1}{2}\Big\{ 1 - e^{- \gamma_{3} t}  \cos^{2} 2 \theta
 - e^{-\gamma t}\,\sin^{2} 2 \theta
 \\
&&
\times \Big[\cosh\lf(\Omega_{\alpha} t \ri) - \frac{\alpha \sin (2 \phi)\, \sinh\lf(\Omega_{\alpha} t \ri)}{  \Omega_{\alpha}  }\Big]\Big\}\,.
\eea
In a similar way, for  anti-neutrino, one has
\bea\non\label{pB}
&& P_{\overline{\nu}_{\sigma} \rightarrow \overline{\nu}_{\varrho}}(t) = \frac{1}{2}\Big\{1 - e^{- \gamma_{3} t } \cos^{2} 2 \theta
-  e^{-\gamma t} \, \sin^{2} 2 \theta
\\
&& \times \Big[\cosh\lf(\Omega_{\alpha} t \ri) +
\frac{\alpha \sin (2 \phi)\, \sinh\lf(\Omega_{\alpha} t  \ri)}{  \Omega_{\alpha}  }\Big]\Big\}\,.
\eea
  Eqs.(\ref{pA}) and (\ref{pB}) show
an asymmetry between the transitions ${\nu}_{\sigma} \rightarrow {\nu}_{\varrho}$ and  $\overline{\nu}_{\sigma} \rightarrow \overline{\nu}_{\varrho}$, i.e.  $P_{\nu_{\sigma} \rightarrow \nu_{\varrho}}(t)  \neq P_{\overline{\nu}_{\sigma} \rightarrow \overline{\nu}_{\varrho}}(t)$.
 Such an asymmetry, is due to Majorana phase $\phi$
and appears also in the probability of an electron, muon or tau neutrino preserving its flavor $\sigma$, ($\sigma = e, \mu, \tau)$, i.e.
 $P_{\nu_{\sigma} \rightarrow \nu_{\sigma}}(t)  \neq P_{\overline{\nu}_{\sigma} \rightarrow \overline{\nu}_{\sigma}}(t)$.
 The $CP$ violation, induced by the oscillation formulas Eqs.(\ref{pA}) and (\ref{pB}) is explicitly  given
by,
\bea\label{CP-viol}\non
\Delta_{CP}^{M}(t) & = & P_{\nu_{\sigma} \rightarrow \nu_{\varrho}}(t)  - P_{\overline{\nu}_{\sigma} \rightarrow \overline{\nu}_{\varrho}}(t)
\\ &  = &   \sin^{2} 2 \theta \,
\frac{\alpha \sin (2 \phi)\, \sinh\lf(\Omega_{\alpha} t  \ri)}{ \Omega_{\alpha}  }\, e^{-\gamma t} \,.
\eea

The definition of the  T-violating quantity in the case of dissipative matter  is more delicate.
Indeed, the decoherence and the dissipation induce an explicit violation of the $T$ symmetry,
which is independent on the nature of the particle.
Here we are interested to the study of $T$ symmetry in neutrino oscillation.

In QM mixing treatment, the T violating asymmetry can be obtained by means of two equivalent definitions, as follows
\bea\non
\Delta_{T}^{M}(t) & = & P_{\nu_{\sigma} \rightarrow \nu_{\varrho}}(t)  - P_{ {\nu}_{\varrho} \rightarrow  {\nu}_{\sigma}}(t)
\\
 &  = &   P_{\nu_{\sigma} \rightarrow \nu_{\varrho}}(t)  - P_{ {\nu}_{\sigma} \rightarrow  {\nu}_{\varrho}}(-t)\,.
\eea
However, in the presence of decoherence, the definition
$\Delta_{T}^{M}(t) =   P_{\nu_{\sigma} \rightarrow \nu_{\varrho}}(t)  - P_{{\nu}_{\sigma} \rightarrow  {\nu}_{\varrho}}(-t)\,, $
cannot be used, since the complete positivity of the matrix density is not   satisfied for any time.
 Indeed, one has
  \bea\label{T1-viol}
 {\Delta_{T}^{M}}(t) &= & P_{\nu_{\sigma} \rightarrow \nu_{\varrho}}(t)  - P_{ {\nu}_{\sigma} \rightarrow  {\nu}_{\varrho}}(- t)
\\ \non
&= & \sin^{2} 2 \theta \Big[ \frac{\alpha \sin (2 \phi)\, \sinh\lf(\Omega_{\alpha} t \ri)
\cosh (\gamma t)}{\Omega_{\alpha}  }
\\  \non
& + & \sinh (\gamma t) \cosh\lf(\Omega_{\alpha} t \ri)
 \Big] + \sinh (\gamma_{3} t) \cos^{2} 2\theta \,.
\eea
The presence of hyperbolic functions in Eq.(\ref{T1-viol}) induces, for sufficiently long time, a violation of the positivity of $\rho$.
This fact produces a value of $ {\Delta_{T}^{M}}$ not included in the interval $ [-1,1] $. Such a result is not physically acceptable.

On the contrary, the relation,
$\Delta_{T}^{M}(t) = P_{\nu_{\sigma} \rightarrow \nu_{\varrho}}(t)  - P_{ {\nu}_{\varrho} \rightarrow  {\nu}_{\sigma}}(t)$,
is  defined properly at any time $t$. By using such a relation, we have,
\bea\label{T2-viol}
\Delta_{T}^{M}(t) = 0\,,
\eea
i.e., the two flavor Majorana neutrino oscillation, in the presence of decoherence, does not violate the $T$ symmetry.

The $CPT$ invariance imposes the relationship $\Delta_{CP}  =  \Delta_{T} $.
However, by comparing Eq.(\ref{CP-viol}) with Eq.(\ref{T2-viol}),
we have $\Delta_{CP}^{M} \neq  \Delta_{T}^{M}$, which implies the violation of the $CPT$ symmetry for Majorana neutrinos, $\Delta_{CPT}^{M} \neq 0$.

Let us consider now  Dirac neutrinos. The phase $\phi$ can be set equal to zero, then  the   oscillation formulas
$P_{\nu_{\sigma} \rightarrow \nu_{\varrho}}(t) $ and $ P_{\overline{\nu}_{\sigma} \rightarrow \overline{\nu}_{\varrho}}(t)$ are equivalent,
  $P_{\nu_{\sigma} \rightarrow \nu_{\varrho}}(t) = P_{\overline{\nu}_{\sigma} \rightarrow \overline{\nu}_{\varrho}}(t)$ and reduce to
\bea\label{pC}\non
P_{\nu_{\sigma} \rightarrow \nu_{\varrho}}(t) & = & P_{\overline{\nu}_{\sigma} \rightarrow \overline{\nu}_{\varrho}}(t)
=
 \frac{1}{2}\Big[1 - e^{- \gamma_{3} t}  \cos^{2}  2 \theta
\\
& - &  e^{-\gamma t} \sin^{2} 2 \theta\, \cosh\lf(\Omega_{\alpha} t \ri)
 \Big]\,.
\eea
In such a  case, the neutrino oscillation  preserves the  $CP$ and $T$ symmetries, $\Delta_{CP}^{D} = \Delta_{T}^{D} = 0$.
The above results show that the decoherence could produce another difference between Majorana and Dirac neutrinos, i.e. the   $CPT$ symmetry is violated by Majorana neutrinos, but it is preserved by Dirac neutrinos.

The kind of $CPT$ violation here presented is due to the mixing in the presence of the decoherence. It could represent an effect induced by the quantum gravity \cite{Rovelli}.
 We emphasize that such a violation is different from an explicit $CPT$ symmetry breaking in the Hamiltonian dynamics such that $[CPT, H] \neq 0$.
 In such a case, a possible cause of the $CPT$ breaking can be represented by the Lorentz violation due to a propagation in a curved space
 violating the Lorentz invariance. In the present  framework, the decoherence  may lead to an effectively ill-defined $CPT$ quantum mechanical operator \cite{Rovelli1}.

 Another result of the present paper, is the discovery of the fact that the oscillation formulas  for Majorana neutrinos depend  on the choice of the mixing matrix $U$.
 Different choices of $U$ lead  to different oscillation formulas. Indeed, Eqs.(\ref{pA}) and (\ref{pB}) are obtained by using the mixing relations given in  Eq.(\ref{U2}).
 On the other hand, by using the mixing matrix of Eq.(\ref{U1}), the oscillation formula for neutrinos Eq.(\ref{pA}) is replaced by the one  for antineutrinos Eq.(\ref{pB}) and viceversa.
The dependence of the oscillation formulas by the choice of the representation of the mixing matrix characterizes the Majorana neutrinos.
Therefore, if the  neutrinos are Majorana particles, the study of the oscillation formulas in long baseline experiments could also allow the determination of the right mixing matrix.

Notice that similar effects are produced by the following dissipator
\bea\label{Dissipator3}
D_{\lambda \mu}\,=\,-
\left(
  \begin{array}{cccc}
    0 & 0 & 0 & 0 \\
    0 & \gamma_{1} & 0 & 0 \\
    0 & 0 & \gamma_{2} & \delta \\
    0 & 0 & \delta & \gamma_{3} \\
  \end{array}
\right).
\eea
  On the other hand, the dissipator
\bea\label{Dissipator4}
D_{\lambda \mu}\,=\,-
\left(
  \begin{array}{cccc}
    0 & 0 & 0 & 0 \\
    0 & \gamma_{1} & 0  & \beta \\
    0 & 0  & \gamma_{2} & 0 \\
    0 & \beta & 0 & \gamma_{3} \\
  \end{array}
\right),
\eea
generates oscillation formulas depending on the phase $\phi$, but there is no $CP$ and $CPT$ violations,
being, in such a case, $P_{\nu_{\sigma} \rightarrow \nu_{\varrho}}(t) = P_{\overline{\nu}_{\sigma} \rightarrow \overline{\nu}_{\varrho}}(t)$.

In order to emphasize the role of off-diagonal elements in the dissipator, we compare the above results with the one obtained  in the case of diagonal dissipator, i.e. $\alpha = 0$, and  $ D_{\mu\nu} \! =\! - \text{diag}(0, \gamma, \gamma,  \gamma_{3}) $.
In such a case, we have   $ \frac{\gamma_3}{2} \leq \gamma $.  The oscillation formulas for Majorana neutrinos are independent on  $\phi$ and coincides with the ones of Dirac neutrinos,
 \bea\non  \label{diagon}\label{diag}
&& P_{\nu_{\sigma} \rightarrow \nu_{\varrho}}(t)   =   P_{\overline{\nu}_{\sigma} \rightarrow \overline{\nu}_{\varrho}}(t)
\\
 && = \frac{1}{2}\Big[1 - e^{- \gamma_{3} t}  \cos^{2} 2 \theta
- \sin^{2} 2 \theta \cos \lf(\Delta\,  t \ri) e^{-\gamma t}\Big]\,.
\eea
%
In such a case, for two flavor neutrino oscillation, one has, $\Delta_{CP} =  \Delta_{T} = \Delta_{CPT} = 0$. The Pontecorvo formulas \cite{Pontecorvo}  are recovered by setting in Eq.(\ref{diag}), $ \gamma = \gamma_3 = 0 $.

The results here presented hold for neutrino propagation in vacuum and also for $\nu_{\mu}\leftrightarrow\nu_{\tau}$ oscillation through the matter of the Earth.
Indeed, the $ \nu_{\mu} \rightarrow \nu_{\tau} $ oscillations and the $CP$ violation in the Earth are practically identical to the ones in vacuum.

In the case of mixing between  $\nu_{e}$ and  $\nu_{\mu}$  through media,
 the Earth is not charge-symmetric, indeed it contains electrons, protons and neutrons, but not contains their antiparticles.
Then, the behavior of neutrinos is different with respect to the one of antineutrinos, also without decoherence (we have to consider $R_-$ in $\Delta m_{m}^{2}$ and $\sin 2 \theta_{m}$ for neutrinos and $R_+$ for antineutrinos). This fact implies that,  the $\nu_{e} \leftrightarrow \nu_{\mu}$  oscillations in matter break the $CP$ and $CPT$ symmetry also in absence of decoherence.
Therefore, the $\nu_{e} \leftrightarrow \nu_{\mu}$  oscillations in matter cannot be used to study the $CPT$ violation induced by the decoherence. Such an analysis, together with the one on  the neutrino nature  can be better done by studying the  $\nu_{\mu}\leftrightarrow\nu_{\tau}$ oscillation in vacuum or through the   Earth,
and  the  $\nu_{e} \leftrightarrow \nu_{\mu}$ oscillation in vacuum.

\section{Numerical analysis}

We now present a numerical analysis of Eqs.(\ref{pA}), (\ref{pB}), (\ref{pC})
in order to study the nature of neutrinos.
Moreover, we consider
Eq.(\ref{CP-viol})  to study the $CP$   and $CPT$ violations in Majorana neutrinos.

We consider the characteristic parameters of the IceCube DeepCore experiment, which detects  neutrino oscillations from atmospheric cosmic rays, over a baseline across the Earth \cite{Aartsen:2017nmd}.
Such an experiment is sensitive at neutrino energies in the range $(6 - 100) GeV$ and it is mainly sensitive to muon neutrinos.
Therefore, we analyze the mixing between $\nu_{\mu}$ and $\nu_{\tau}$ and we compute the probability of transition $P_{\nu_{\mu} \rightarrow \nu_{\tau}}(x)$ and the corresponding oscillation formula for the anti-neutrinos  $P_{\overline{\nu}_{\mu} \rightarrow \overline{\nu}_{\tau}}(x)$.
We also study the mixing between $\nu_{e}$ and $\nu_{\mu}$ and the oscillation formulas $P_{\nu_{e} \rightarrow \nu_{\mu}}(x)$ and $P_{\overline{\nu}_{e} \rightarrow \overline{\nu}_{\mu}}(x)$.
The range of neutrino energy analyzed for $\nu_{e} \leftrightarrow \nu_{\mu}$ oscillation  is $(0.3 - 5) GeV$, which is characteristic of DUNE experiment.

We use, in natural units, the approximation, $x \approx t$, where $x$ is the distance travelled by neutrinos.
We analyze the  neutrino propagation in the vacuum and through the matter.

In Fig.1, we plot  the oscillation formulas in vacuum and through  Earth, $P_{\nu_{\mu} \rightarrow \nu_{\tau}} $  and $P_{\overline{\nu}_{\mu} \rightarrow \overline{\nu}_{\tau}} $, as a function of the neutrino energy $E$. The plots refer to    Majorana neutrinos and to Dirac neutrinos, (cfr. Eqs.(\ref{pA}), (\ref{pB}), (\ref{pC}), respectively). The comparison with the oscillation formula for  diagonal dissapator, $\alpha = 0$ (cfr. Eq.(\ref{diagon})) and the Pontecorvo-Bilenky oscillation formula    is also analyzed.
In the inset, we plot the value of the $CP$ asymmetry $\Delta_{CP}^{M  }(x) = P_{\nu_{\mu} \rightarrow \nu_{\tau}}(x)  - P_{\overline{\nu}_{\mu} \rightarrow \overline{\nu}_{\tau}}(x)$   as a function of $E$ for the same values of the parameters used in the main plot.
The plots are derived by assuming $\phi = \frac{\pi}{4}$. We used  a distance  equals to   Earth diameter $ x = 1.3\times 10^{4} km $,   considered the energy interval $[6 - 120] GeV$ and the following values of the parameters:  $\sin^{2} \theta_{23} = 0.51 $, $\Delta m^{2}_{23} = 2.55 \times 10^{- 3} eV^{2} $, $\gamma = 4 \times 10^{-24}GeV$, $\gamma_{3} = 7.9 \times 10^{-24}GeV $, $\alpha = 3.8 \times 10^{-24}GeV$ \cite{Coloma}.

\begin{figure}[t]
\begin{picture}(300,180)(0,0)
\put(10,20){\resizebox{8.0 cm}{!}{\includegraphics{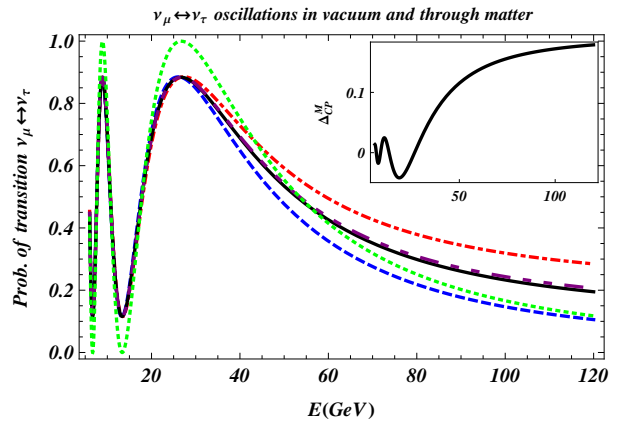}}}
\end{picture}
\caption{\em (Color online) Plots of the oscillation formulas  $P_{\nu_{\mu} \rightarrow \nu_{\tau}} $ (the red dot dashed line) and $P_{\overline{\nu}_{\tau} \rightarrow \overline{\nu}_{\mu}} $ (the blue dashed line) for Majorana neutrinos and for Dirac neutrinos ($\phi = 0$, the black line), as a function of the energy $E$, in vacuum and through matter. The purple,  dashed line is obtained by setting $\alpha = 0$. In this case $P_{\nu_{\mu} \rightarrow \nu_{\tau}}  = P_{\overline{\nu}_{\mu} \rightarrow \overline{\nu}_{\tau}} $ and one has the same formula for Majorana and for Dirac neutrinos. The Pontecorvo formula is represented by the green dotted line. We assume  $\phi = \frac{\pi}{4}$,  $ x = 1.3\times 10^{4} km $ and we use the following experimental values of the parameters:  $\sin^{2} \theta_{23} = 0.51 $, $\Delta m^{2}_{23} = 2.5 \times 10^{- 3} eV^{2} $. Moreover, we set $\gamma = 4 \times 10^{-24}GeV$, $\gamma_{3} = 7.9 \times 10^{-24}GeV $, $\alpha = 3.8 \times 10^{-24}GeV. $
Picture in the inset: plot of $\Delta_{CP}^{M  }(x) $   for the same values of the parameters used in the main plots.
Such plots describe the propagation in vacuum and through   Earth.}
\label{pdf}
\end{figure}
\begin{figure}[t]
\begin{picture}(300,180)(0,0)
\put(10,20){\resizebox{8.0 cm}{!}{\includegraphics{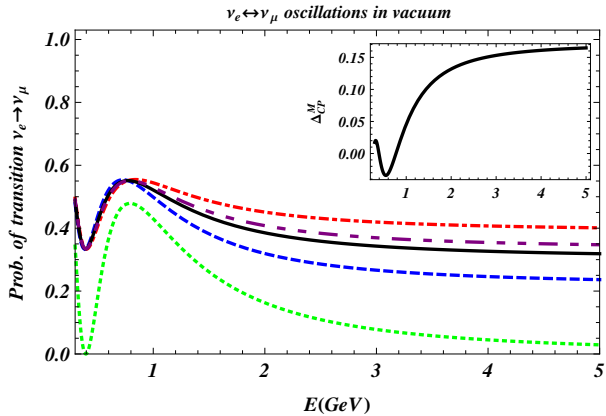}}}
\end{picture}
\caption{\em (Color online) Plots of   $P_{\nu_{e} \rightarrow \nu_{\mu}} $ (red dot dashed line) and $P_{\overline{\nu}_{e} \rightarrow \overline{\nu}_{\mu}} $ (blue dashed line) for Majorana neutrinos and for Dirac neutrinos ($\phi = 0$,  black line), as a function of  $E$, in vacuum. The purple,  dashed line is obtained by setting $\alpha = 0$. In this case $P_{\nu_{e} \rightarrow \nu_{\mu}}  = P_{\overline{\nu}_{e} \rightarrow \overline{\nu}_{\mu}} $. The Pontecorvo formula is represented by the green dotted line. We use the same values of  $\phi  $ and $ x  $ of Fig.(1), moreover, we consider:  $\sin^{2} \theta_{12} = 0.861 $, $\Delta m^{2}_{ 12} = 7.56 \times 10^{- 5} eV^{2} $, $\gamma = 1.2 \times 10^{-23 }GeV$, $\gamma_{3} = 2.3 \times 10^{-23}GeV $, $\alpha = 1.1 \times 10^{-23}GeV$.
Picture in the inset:  plot of $\Delta_{CP}^{M  }(x) $.}
\label{pdf}
\end{figure}

In Fig.2, we plot the oscillation formulas in vacuum, $P_{\nu_{e} \rightarrow \nu_{\mu}} $  and $P_{\overline{\nu}_{e} \rightarrow \overline{\nu}_{\mu}} $ and in the inset   the $CP$ asymmetry $\Delta_{CP}^{M  } = P_{\nu_{e} \rightarrow \nu_{\mu}}(t)  - P_{\overline{\nu}_{e} \rightarrow \overline{\nu}_{\mu}}(t)$ . We use the same values of $\phi$ and $x$ considered in Fig.1, moreover we use  $\sin^{2} \theta_{12} = 0.861 $, $\Delta m^{2}_{ 12} = 7.56 \times 10^{- 5} eV^{2} $, $\gamma = 1.2 \times 10^{-23 }GeV$, $\gamma_{3} = 2.23 \times 10^{-23}GeV $, $\alpha = 1.1 \times 10^{-23}GeV$ \cite{Balieiro}.

By analyzing the plots in Figs.1 and 2,  one  can see  that the differences between Majorana and Dirac neutrinos, the $CP$  and $CPT$ violations are, in principle,  detectable.
Indeed, considering the $CP$ violation, which, in the case of two flavor neutrino mixing, is different from zero only for Majorana neutrinos, one   finds:
a) for $\nu_\mu \leftrightarrow \nu_\tau$ neutrino oscillation in vacuum and through matter, in particular ranges of the energy,
$\Delta_{CP}^{M (\mu-\tau)} \sim 0.18  $,
b) for $\nu_e \leftrightarrow \nu_\mu$ neutrino oscillation in vacuum,
$\Delta_{CP}^{M (e-\mu)} \sim 0.16  $ (see Figs. 1 and 2).

In Fig.3, we include the matter effect for the oscillation $ \nu_{e}  \leftrightarrow \nu_{\mu} $ and we plot the $CP$ asymmetry $\Delta_{CP}  = P_{\nu_{e} \rightarrow \nu_{\mu}}(t)  - P_{\overline{\nu}_{e} \rightarrow \overline{\nu}_{\mu}}(t)$, for Majorana and for Dirac neutrinos in the presence of decoherence with off-diagonal term. Moreover, we plot  $\Delta_{CP}$ in absence of decoherence.
 In our computations, we consider electron number density $n_{e} = 2.36   cm^{-3} N_A  $  and the range of energy $[0.3  - 1] GeV$.
The value of $n_{e}$ represents the weighted arithmetic mean of the mean electron densities in the Earth mantle $n_{e}^{m}$ and in the Earth core  $n_{e}^{c}$.
   (The Earth mantle has a radius of $R_m = 2885 km$ and an estimated mean electron number density $n_{e}^{m} \simeq 2.2 cm^{-3} N_A  $, the Earth core has a radius of $R_c = 3846 km$ and an estimated mean electron number density $n_{e}^{c} \simeq 5.4 cm^{-3} N_A  $).
The plots show different behaviors of $\Delta_{CP}$ for Majorana and for Dirac neutrinos in the presence of decoherence with off-diagonal term.
Such behaviors are different with respect to the one of $\Delta_{CP}$ obtained by considering the two flavor neutrino mixing without decoherence.

%

\begin{figure}[t]
\begin{picture}(300,180)(0,0)
\put(10,20){\resizebox{8.0 cm}{!}{\includegraphics{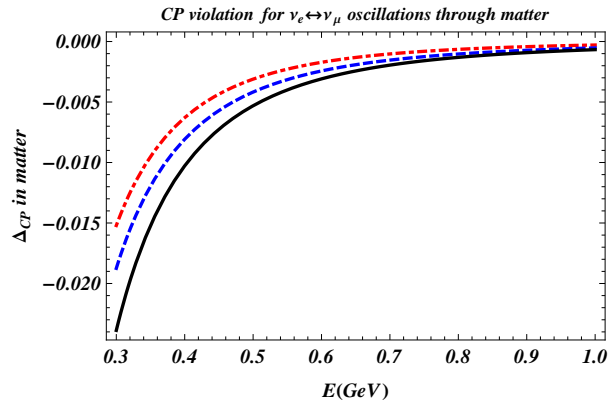}}}
\end{picture}
\caption{\em (Color online) Plots of $\Delta_{CP}  = P_{\nu_{e} \rightarrow \nu_{\mu}}(t)  - P_{\overline{\nu}_{e} \rightarrow \overline{\nu}_{\mu}}(t)$, through matter, for Majorana neutrinos (the red dot dashed line), for Dirac neutrinos (the blue dashed line) in the presence of decoherence with off-diagonal term and for neutrinos in absence of decoherence (the black dotted line).  In the plots, we consider the same parameters of Fig.(2) and the energy interval $E \in (0.3 - 1)GeV $.}
\label{pdf}
\end{figure}

\section{Conclusions}

We have studied different features of the phenomenon of the decoherence   in neutrino oscillations.
We have shown the possible $CPT$ symmetry breaking in the Majorana neutrino oscillation, we have shown that the probability of transitions for Majorana neutrinos   depend  on the representation of the  mixing matrix, and
 we have study  the phenomenological differences between Majorana and Dirac neutrinos in their oscillations.

 By using the data of   IceCube DeepCore and   DUNE experiments, and by considering the constraints on decoherence parameters \cite{Coloma,Balieiro}, we have analyzed the oscillation formulas for atmospheric neutrinos, $P_{\nu_{\mu} \rightarrow \nu_{\tau}} $  and $P_{\overline{\nu}_{\mu} \rightarrow \overline{\tau}_{\mu}} $, and for neutrinos produced in accelerator, $P_{\nu_{e} \rightarrow \nu_{\mu}} $  and $P_{\overline{\nu}_{e} \rightarrow \overline{\nu}_{\mu}} $.
We have studied the behaviors of   $CP$ and  $CPT$ violations in neutrino oscillation and
we  have shown that, the differences between Majorana and Dirac neutrinos, together with the $CPT$ violation could be detected
 if the phenomenon of decoherence is taken into account during the neutrino propagation in long baseline experiments.
 Moreover, the oscillation formulas could provide a tool to determine the choice of the mixing matrix for Majorana neutrinos, if the neutrino is a Majorana fermion.

Since, at the Earth densities, the MSW effect affects only the $ \nu_{e} \leftrightarrow \nu_{\mu} $ oscillations  and such oscillations in matter are neither $CP$, nor $CPT$ invariant, thus, $ \nu_{e} \leftrightarrow \nu_{\mu} $ oscillations in matter  are not suitable  to study the possibility of $CPT$ breaking induced by the decoherence. For such an analysis, one has to consider long baseline experiments analyzing atmospheric neutrinos, such as   the  IceCube experiment, or one has to study the $ \nu_{e} \leftrightarrow \nu_{\mu} $ oscillations in vacuum.
We point out that the $CPT$ violation, the difference between the oscillation formulas of Majorana and Dirac neutrinos, and the dependence of such formulas on the representation of the mixing matrix appear only in the cases of a not diagonal form of the dissipator, similar to the one presented in Eq.(\ref{Dissipator2}). In the case of diagonal dissipator, such effects disappear.
Then experiments like IceCube, could   allow the determination of the correct form of the matrix describing the decoherence, if such phenomenon is relevant in neutrino oscillation.

Neutrino decoherence  and CPT violation could be signals of quantum gravity.
 Therefore, our analysis could open new interesting scenarios not only in the study of neutrinos, but also in other fields of fundamental physics.

Notice also that,  non-perturbative field theoretical
effects of particle mixing \cite{Blasone:1998hf}, \cite{Capolupo:2006et}, can be neglected
in the our treatment.

\section*{Acknowledgements}
A.C. and G.L. acknowledge partial financial support from MIUR and INFN and the COST Action CA1511 Cosmology
and Astrophysics Network for Theoretical Advances and Training Actions (CANTATA).
S.M.G. acknowledge support by the H2020 CSA Twinning project No. 692194, "RBI-T-WINNING''.


\begin{thebibliography}{99}
\bibitem{note}
Indeed,  Dirac fields $\psi$ are composed by the left-handed $ \psi_{L}$ and the right-handed $\psi_{R}$ components. Then, in the free Lagrangian $L =  \overline{\psi}(i \gamma^{\mu} \partial_{\mu} - m )\psi $, the mass term can be written as
$L_{m} = - m \overline{\psi}\psi = - m (\overline{\psi_{L}+\psi_{R}})(\psi_{L}+\psi_{R}) = - m(\overline{\psi_{L}}\psi_{R} + \overline{\psi_{R}}\psi_{L})$, (the $\overline{\psi_{L}}\psi_{L}$ and
$\overline{\psi_{R}}\psi_{R}$ terms vanish identically). In such a case, $L_{m}$ (and then $L$) is invariant under the $U(1)$ transformations $\psi \rightarrow e^{i \phi} \psi$ and $\overline{\psi} \rightarrow e^{- i \phi} \overline{\psi}$. Therefore the charges (electric, leptonic, etc.) associated with the transformations are conserved.
On the other hand, for Majorana fields the right-handed component of $\psi$ is the charge conjugate of its left-handed component, $\psi_{R} = (\psi_{L})^{c}$, where $\psi^{c}  = C \overline{\psi}^{T}$, and $C$ is the charge conjugation matrix.  Then one has $\psi = \psi_{L} + (\psi_{L})^{c} $. In such a case, the mass lagrangian is
$L_{m} = - \frac{m}{2} \lf[ \overline{(\psi_{L})^{c}}\psi_{L} + \overline{\psi_{L}}(\psi_{L})^{c}  \ri]
= \frac{m}{2} \lf[ \psi_{L}^{T} C^{-1}\psi_{L} + h.c. \ri]$, (being $\overline{\psi^{c}} = - \psi^{T} C^{-1}$), i.e. $L_{m}$ has the structure
$\psi_{L} \psi_{L} + h.c.$,
which therefore, under the $U(1)$ transformations, breaks all the $U(1)$-charges of two units.
\end{thebibliography}
\end{document}